\documentstyle[epsf]{laa}
\begin{document}
\newcommand{\teff}  	{$T_{\mathrm{eff}}$}	
\renewcommand{\lg}  	{$\log g$}			
\newcommand{\mh}    	{[M/H]}			
\newcommand{\feh}    	{[Fe/H]}		

\newcommand{\al}    {\em et al.}
\newcommand{\eg}    {\em e.g.}
\newcommand{\ie}    {\em i.e.}

\newcommand{\aap}  [2]{A\&A {\bf #1}, \rm #2}
\newcommand{\aas}  [2]{A\&A Suppl. {\bf #1}, \rm  #2}
\newcommand{\aj}   [2]{AJ {\bf #1}, \rm #2}
\newcommand{\apj}  [2]{Ap. J. {\bf #1}, \rm #2}
\newcommand{\apjl} [2]{Ap. J. Letter {\bf #1}, \rm #2}
\newcommand{\apjs} [2]{Ap. J. Suppl. {\bf #1}, \rm #2}
\newcommand{\araa} [2]{A\&ARA {\bf #1}, \rm #2}
\newcommand{\pasp} [2]{PASP {\bf #1}, \rm #2}
\newcommand{\mnras}[2]{MNRAS {\bf #1}, \rm #2}

\thesaurus{...}

\title{A standard stellar library for evolutionary synthesis}
\subtitle{II. The M dwarf extension}

\author{T. Lejeune    \inst{1,2}
 \and F. Cuisinier \inst{3} \and R. Buser \inst{1} }

\institute{Astronomisches Institut der Universit\"at Basel,
           Venusstr. 7, CH--4102 Binningen, Switzerland
     \and  Observatoire de Strasbourg, 11, rue de l'Universit\'e,
           F--67000 Strasbourg, France
     \and  Universidade de S\~ao Paulo, IAG,
	   Dept. de Astronomia, CP 9638, 
           S\~ao Paulo 01065-970, Brasil}

\offprints{T. Lejeune: \\
lejeune@astro.unibas.ch}
\maketitle

\begin{abstract}

A  standard library   of theoretical   stellar  spectra intended   for
multiple   synthetic   photometry applications  including     spectral
evolutionary synthesis is presented.  The  grid includes M dwarf model
spectra, hence complementing  the first library version established in
Paper I (Lejeune, Cuisinier \& Buser 1997).   It covers wide ranges of
fundamental parameters: {\teff} : 50,000 K  $\sim$ 2000 K, {\lg} : 5.5
$\sim$ -1.02, and {\mh} : +1.0 $\sim$ -5.0.  A correction procedure is
also   applied  to the   theoretical   spectra  in  order to   provide
color-calibrated flux distributions over  a large domain of  effective
temperatures.  For this purpose, empirical {\teff}--color calibrations
are constructed between 11500 K  and 2000 K, and {\em  semi}-empirical
calibrations for  non-solar  abundances  ({\mh}  = -3.5  to  +1.0) are
established.  Model  colors  and bolometric corrections   for both the
original    and   the   corrected   spectra,    synthesized   in   the
$(UBV)_{J}(RI)_{C}JHKLL'M$ system,  are given  for  the full  range of
stellar parameters.  We find that the corrected spectra provide a more
realistic  representation   of empirical   stellar colors,  though the
method  employed is not  completely adapted  to the lowest temperature
models.   In particular the original  differential  colors of the grid
implied by metallicity   and/or luminosity changes are  not  preserved
below 2500 K.    Limitations of the correction   method used are  also
discussed.
\end{abstract}

\section{Introduction}

Grids of theoretical stellar spectra providing  models at low and high
metallicities are  indispensable for modelling  the chemical evolution
of the integrated light of stellar systems from theoretical isochrones
used to describe  the time-dependent stellar population. However, even
the  combined    uses  of   available modern  stellar    libraries  in
evolutionary synthesis   studies ({\eg}   Buzzoni 1989, Worthey  1994,
Bressan {\al} 1994) have been handicapped by intrinsic inhomogeneities
and incompleteness. Furthermore, one of  the most serious difficulties
arises from the fact that the synthetic spectra and colors provided by
most theoretical libraries still  show large systematic  discrepancies
with calibrations based on spectroscopic and photometric observations.
This  is particularly true at  low effective temperatures for which an
accurate modelling of the stellar spectra requires important molecular
opacity data which are not yet completely available. Ultimately, these
limitations   lead unavoidably   to   serious   uncertainties in   the
interpretation of the population model.

In order to overcome these major shortcomings,  we have undertaken the
construction of a   comprehensive combined library of  {\em realistic}
stellar flux  distributions intended for  population and  evolutionary
synthesis studies.   A  preliminary version of   such a {\em standard}
grid was   presented in  Lejeune  {\al}  (1997, Paper I,   referred to
hereafter as LCB97), along with an algorithm developed for correction
and calibration of the (theoretical)  spectra.  In this previous grid,
M   dwarf   models, which  are   important   for the determination  of
mass-to-light ratios in stellar populations, were missing.  We present
here  a  more  comprehensive  library which  incorporates  these dwarf
spectra.  The construction of this basic combined library is presented
in the following  section.  The empirical {\teff}--color relations  in
UBVRIJHKL photometry, required to calibrate the spectra, are presented
in section
\ref{sect:tempcol_cal}.  Section \ref{sect:calib_spec} is dedicated to
the  correction of  the  theoretical spectra;  in  particular, we also
examine the properties  and the limitations  of  the method used  when
applied to   the  M dwarf  models.   Finally,  we summarize  the  main
properties  of  this new standard   library  in view of  its synthetic
photometry applications.

\section{The model library}
\label{sect:mod_lib}

\subsection{Construction of a combined library}

\begin{figure}
\epsfxsize=9cm
\epsffile{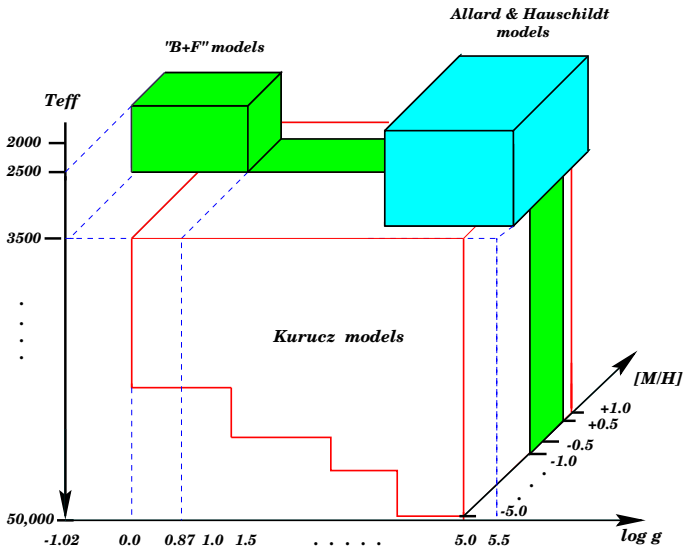}
 \caption  []{Coverage of the  final  combined library  in the stellar
 parameter space. ``B+F'' models  are those constructed for M  giants
 from Fluks {\al} and BBSW spectra (see LCB97).}
\label{f:bcl96.3D}
\end{figure}

Proceeding with  the work undertaken in   LCB97, we have  built a more
extensive library  providing   now almost  complete  coverage  of  the
stellar  parameter ranges in {\teff}, {\lg},  {\mh} which are required
for  population and evolutionary synthesis  studies.  This new grid is
complementing the  previous version by the  addition of  M dwarf model
spectra.  Different basic libraries have   been assembled: the  Kurucz
(1995, [K95]) models provide wide coverage in {\teff} (50,000 K $\sim$
3500 K), {\lg} (5.0 $\sim$ 0.0), and {\mh} (+1.0 $\sim$ -5.0), whereas
in the temperature range 3500 K $\sim$ 2500 K the M giants spectra are
represented  by the  hybrid  {\em   ``B+F''} models constructed   from
spectra of Fluks  {\al} (1994) and Bessell  {\al} (1989, 1991  [BBSW])
(LCB97).  For the M dwarf  models, we introduced the synthetic spectra
of cool stars originating from the Allard \& Hauschildt (1995, [AH95])
grid.  Thus, the  previous  library is extended  by  models within the
following parameter ranges:   3500 K $>$ {\teff}  $\geq$  2000  K, 5.5
$\geq$ {\lg} $\geq$ 3.5 and +0.5 $\geq$ {\mh} $\geq$ -4.0.  Fig.
\ref{f:bcl96.3D}  gives a 3-D   representation   of the new   combined
library.

The  ``base  model grid'' of    AH95  atmosphere models  (referred  to
hereafter as the {\em ``Extended''}  models) used here present several
improvements compared  to  the previous  generation  models  of Allard
(1990).  In particular, new molecular opacities have been incorporated
and the  opacity sampling  technique has  been introduced in  order to
improve the treatment of some of the  atomic and molecular lines.  For
solar metallicity,   we used an upgraded  version  (referred to as the
{\em  ``NextGen''} models version) of  the   ``base model grid''.   In
these new  models  a  more extensive   list of  12  million  TiO lines
(J{\o}rgensen 1994) has  also  been  included, with a  more  rigourous
line-by-line   treatment   instead  of  the   Just  Overlapping   Line
Approximation  (JOLA) previously used.      The  main effect   is  the
reduction of the derived effective temperature by  $\sim$ 150 K (Jones
{\al}  1996).   We also  found  that these  new   models provide  more
realistic  UBV  colors  than  the  previous  generations  (see Section
\ref{sect:corr_spec}).  Because  the  {\em ``NextGen''} model grid  is
still incomplete,  only  solar-abundance spectra  can   be included in
order to  cover the whole  range of {\teff}  and {\lg} provided by the
hybrid   library.  For other  metallicities,   we  still use the  {\em
``Extended''} models.

In the combined library, all the original spectra were rebinned on the
same  K95 wavelength  grid, from  9.1 nm to   160 $\mu$m, with a  mean
resolution of  $\sim$ 10 {\AA} in  the ultraviolet and $\sim$ 20 {\AA}
in the visible.  Recall  that the M  giant  model spectra ``B+F''  are
metallicity-independent   blueward  of  4900    {\AA}  owing   to  the
introduction of the  Fluks {\al}  models  in this spectral range,  and
that a  black-body tail was attached for  $\lambda \geq$ 4070  nm (see
LCB97).  The AH95 spectra stop at  20 $\mu$m, and  for $\lambda$ $>$ 3
$\mu$m the resolution is not sufficient for an accurate description of
the synthetic spectrum.  In  order to cover  the whole K95  wavelength
range, we then connected a  black-body to the synthetic spectra beyond
$\sim$ 5 $\mu$m (M band).

These  model spectra  -- except for  the  updated {\em NextGen} models
used  in  this   work for  solar metallicity  --   are  available on a
CD-ROM\footnote{The previous version of the  combined library used the
{\em  ``Extended''} M dwarf models  of Allard \& Hauschildt (1995) for
solar metallicity}  collecting various materials for  galaxy evolution
modelling (Leitherer {\al} 1996), or  by request on anonymous ftp, and
have been  used for single stellar   populations models (Bruzual 1996,
Bruzual {\al} 1997, Lejeune 1997).

\subsection{Model colors and bolometric corrections}

Synthetic UBVRIJHKLL'M  colors  and  bolometric corrections have  also
been computed from the theoretical stellar energy distributions (SEDs)
over the whole  range of parameters available in  the grid,  using the
filter transmission functions given by Buser (1978) for $UBV$, Bessell
(1979) for $(RI)_{C}$, and Bessell \& Brett (1988) for $JHKLL'M$.  The
zero-points were defined  from the {\em Vega}  model spectra of Kurucz
(1991) by fitting  to the observed  values from Johnson (1966) (U-B  =
B-V = 0.0), Bessell \&  Brett (1988) (J-H  = H-K =  K-L = K-L' = K-M =
0.0), and Bessell  (1983)  (V-I  =  0.005, R-I  =  -0.004).   For  the
bolometric    corrections,  we followed    the zero-pointing procedure
described in  Buser \& Kurucz (1978).   We first (arbitrarily) set the
smallest bolometric correction for the ({\teff} = 7000 K, {\lg} = 1.0)
model to  {\em zero}.  This produces   $BC'_{\mathrm{V}}$ = -0.190 for
the  solar model, {\teff}    = 5777 K, {\lg}   =  4.44 (LCB97).    The
zero-point is then defined,
\begin{equation}
BC_{\mathrm{V}} = BC'_{\mathrm{V}} + 0.082 \, ,
\label{eq:bcv}
\end{equation}
in order for the present theoretical  calculations to provide the best
fit of  the  empirical bolometric-correction scale of   Flower (1996).
Adopting these definitions and the standard  value of the solar radius
(Allen 1973) we find  $L = 3.845   10^{33}$ erg/s, $M_{\mathrm{V}}$  =
4.854 and $BC_{V}    = -0.108$  for  the   solar model.   Tables    of
theoretical colors and bolometric corrections  for the whole range  of
stellar parameters in the grid are available in electronic form.

\section{Temperature--color calibrations}
\label{sect:tempcol_cal}

\subsection{Empirical calibrations at solar metallicity}
\label{sect:emp_cal}

In   the correction   procedure   defined  in   LCB97, the   empirical
temperature--color calibrations are the  basic links between the model
colors and  the observations.  First, they provide  a crucial point of
comparison, and  secondly, they are used  to define the  empirical and
theoretical pseudo-continua from which  the correction  functions will
be defined in the following.   Over a large  range of temperatures, we
used   the     empirical   {\teff}--$(UBV)_{J}(RI)_{C}JHKL$  relations
discussed in  detail in  Paper I, but   the inclusion of the  M  dwarf
models in the library now leads us to extend the previous calibrations
to the bottom  of the  main sequence.   Observations of very  low-mass
stars being still    rather  fragmentary, the construction    of these
temperature--color  relations require indirect empirical methods which
are now described.

\subsubsection{Dwarfs}

Over the temperature range  11500 K $\geq$ {\teff}  $\geq$ 4250 K, the
empirical temperature--color sequences are  based upon the temperature
scale of Schmidt-Kaler (1982) for U--B and Flower (1996) for B--V, and
the two-color relations compiled  in the literature (FitzGerald  1970,
Bessell 1979, and Bessell \& Brett 1988).

For   {\teff} $\leq$   4000   K, the    temperature  scale   is   very
controversial, in particular because of  the difficulty to  accurately
model the complex featured  M dwarf spectra.  Due to  the lack of very
reliable model-atmospheres,  indirect  methods such   as blackbody  or
gray-body fitting  techniques  have been  used to  estimate  effective
temperatures   of  the  intrinsically   faintest  stars  (Veeder 1974,
Berriman  \& Reid 1987, Bessell  1991, Berriman, Reid \& Leggett 1992,
Tinney, Mould  \& Reid 1993).  In  practice, the  temperatures derived
from  fitting  to model spectra  ({\eg}  Kirkpatrick {\al} 1993, Jones
{\al}  1994)  are  systematically  $\sim$   300  K warmer than   those
estimated by  empirical methods.  Recently,  a  redetermination of the
effective temperatures  using the  {\em  NextGen} version of the  AH95
model spectra has been  proposed.  Leggett {\al} (1996)  used observed
infrared low-resolution spectra and photometry to compare with models.
They found radii and effective  temperatures which are consistent with
estimates based only  on  photometric  data.  Their study  shows  that
these updated  models should provide, for the  first time, a realistic
temperature   scale for  M dwarfs.  On   the other  hand, Jones  {\al}
(1996), using a specific spectral  region (1.16--1.22 $\mu$m) which is
very  sensitive to parameter changes  of  M dwarfs (Jones {\al} 1994),
have derived stellar  parameters  by fitting synthetic  spectra  for a
limited sample of well-known low-mass stars.  They  found that the new
models provide   reasonable representations  of the   overall spectral
features, with   realistic relative  strength  variations  induced  by
changes in stellar parameters.

\begin{figure}
\epsfxsize=9cm
\epsffile {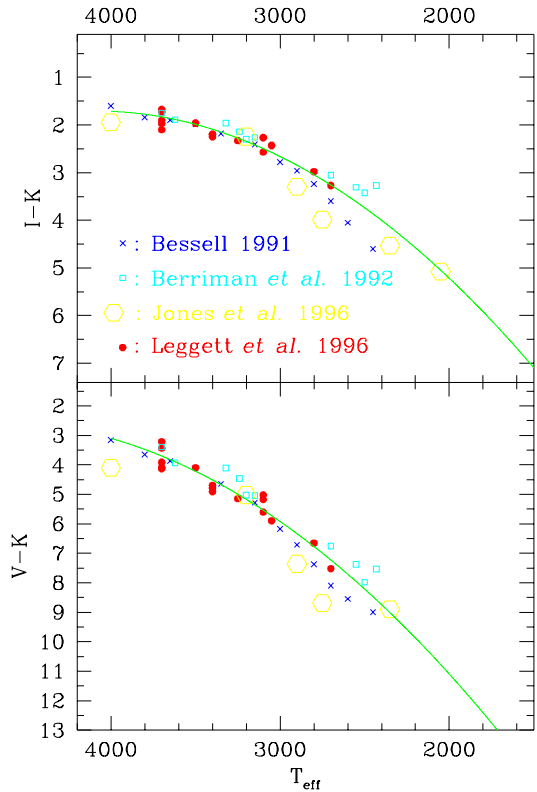}
 \caption []{A comparison of different  temperature scales of M dwarfs
 adopted   by  several authors. The  solid  line  is a  polynomial fit
 performed on all  the data, except those of  Jones {\al} 1996  (large
 open symbols). See text for explanations.}
\label{f:vk_teff.Mdw}
\end{figure}

Based on these promising -- although preliminary -- results of Leggett
{\al},   providing  closer    agreement    between  theoretical    and
observational temperature   scales,   we   adopted  a  mean   relation
constructed from a compilation  of   the results of Bessell    (1991),
Berriman   {\al}   (1992),     and  Leggett  {\al}     (1996).    Fig.
\ref{f:vk_teff.Mdw} shows  the different effective  temperature scales
adopted by these authors, as a function of  I--K and V--K.  All the IR
photometric data have been  transformed to the homogeneous JHKL system
of Bessell \& Brett.  The solid line is a  polynomial fit derived from
these data.    For comparison, we  have  also added in the  figure the
{\teff}--values  estimated by Jones  {\al} (1996)  for  6 stars (large
open symbols) having VIK photometry data given  in Jones {\al} (1994).
Due to the limited spectral range used,  the temperatures estimated by
Jones {\al} (1996)  are still uncertain, and  hence have not been used
to define the mean relation given in Fig.
\ref{f:vk_teff.Mdw}.  For stars  cooler than 3000 K, the discrepancies
between the  different  temperature    scales appear  slightly    more
pronounced in I--K than in V--K.  For this  reason, V--K was preferred
to I--K for  establishing the mean temperature scale  of M dwarfs.  We
found:

\begin{equation} 
V-K  =        1.18    \:       10^{-6}*(T_{\mathrm{eff}})^{2}       -
0.01*T_{\mathrm{eff}}+ 28.49 \, .
\end{equation}

For practical purposes,  a good approximation  of the inverse relation
is given by:

\begin{equation} 
T_{\mathrm{eff}} = 18.27*(V-K)^{2} - 504.88*(V-K) + 5415 \, .
\end{equation}

Notice that this (V--K)--{\teff} scale perfectly matches the   Bessell
(1991)  calibration above 3000  K, and that the temperatures estimated
by Leggett  {\al} are  also in general   agreement with this relation.
The mean (V--K)--{\teff}  relation defined above  was thus  adopted as
the basic scale for M dwarfs over the range 4000 K $\sim$ 2000 K.

For   {\teff}  between  4000  K    and 2600 K,     we   then used  the
$(BV)_{J}(RI)_{C}JHK$  photometric  data given  in Bessell (1991)
and for  stars   hotter than  $\sim$ 3000    K the  U--B  colors  from
FitzGerald (1970) in order  to relate  V--K to the  other colors  at a
given temperature, via color--color transformations.

\begin{figure}
\epsfxsize=9cm
\epsffile {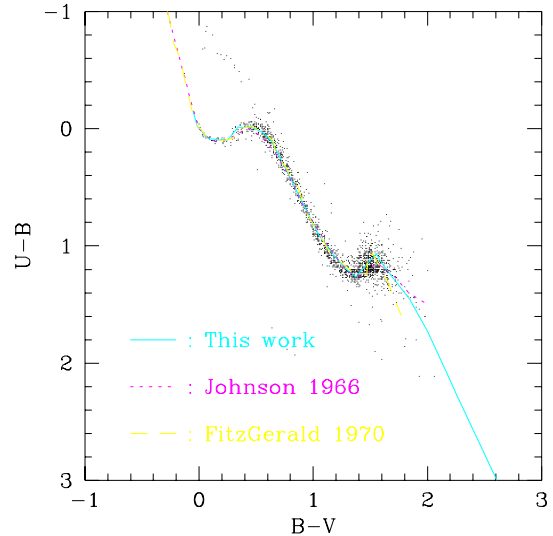}
 \caption []{U--B/B--V empirical sequences  compared to the Gliese  \&
 Jahreiss (1991, CSN3) catalog of nearby stars.}
\label{f:ub_bv.Mdw}
\end{figure}

In order to establish the calibrations down to 2000 K, we used the two
objects  LHS2924 and GD165B with  effective  temperatures estimated by
Jones {\al} 1996  (2350 K and  2050 K, respectively), for which  $VIJHKL$ photometry is  given in Jones  {\al} (1994).  The  {\teff} of
LHS2924 was found  to be in very  good agreement with  our temperature
scale (Fig. \ref{f:vk_teff.Mdw}) whereas  GD165B is the coolest object
with available infrared photometry.  Therefore, the  cool tails of our
empirical calibrations for  the J--H, J--K,  H--K and K--L colors were
required to match these two extreme points.  Nevertheless, since these
two  objects are    good  brown  dwarf  candidates, with   potentially
non-solar abundances   ([Fe/H] $\sim$ -0.5   for LHS2924 and  +0.5 for
GD165B, Jones {\al} 1996), the empirical infrared stellar colors below
2300  K  should  be considered  with  caution.   Furthermore, reliably
accurate UBVRI photometry  data for M  dwarfs cooler than 3000  K (B-V
$\sim$ 1.8) are also  difficult to  obtain  -- and are sparse  indeed.
Some  relevant data can  be found  in the   Gliese \&  Jahreiss (1991)
catalog of nearby stars.  Consequently, we extrapolated down to 2000 K
the  {\teff}--UBVRI   calibrations  defined  above  for  hotter stars.
Compared to the calibrations of  Johnson (1966) and FitzGerald (1970),
our (U--B)-(B--V) relation  yields a better match  of the  extreme red
points   found     in     the     Gliese   \&     Jahreiss     catalog
(Fig. \ref{f:ub_bv.Mdw}).

As before, the surface gravities along the dwarf sequence were defined
at each {\teff} from a ZAMS computed by the  Bruzual \& Charlot (1996,
{\em private communication}) isochrone synthesis program.

\subsubsection{Giants}

For cool giants, the Ridgway {\al} (1980) (V--K)--{\teff} relation was
used as the  basic  temperature scale.  As  in Paper  I, the different
{\teff}--color sequences were constructed by compiling the photometric
calibrations from Johnson  (1966),  Bessell (1979), Bessell   \& Brett
(1988),  and  recent observations of  M   giants given  by Fluks {\al}
(1994).  A 1 $M_{\odot}$  evolutionary track (Schaller {\al} 1992) was
used  to  define the  {\lg}  of the   red giants  with {\teff}  in the
interval 4500 K $\sim$ 2500 K.

\subsection{Semi-empirical calibrations for non-solar abundances}
\label{sect:semi_emp_cal}

In LCB97,  one of the basic assumptions  made to define the correction
process of  the  spectra  is that   the original  model grids  provide
realistic color differences with  respect to metal-content variations.
These  properties,  established in particular  for  the  K95 grid from
Washington  photometry by Lejeune  \& Buser  (1996), can be  naturally
applied to define {\em  semi-empirical} {\teff}--color calibrations at
different metallicities.  This is achieved  by calculating at a  given
effective temperature the  theoretical color differences due  uniquely
to a  change   in  the chemical  composition   (hereafter  called {\em
differential colors}).  These differences are then added to the colors
provided  by  the   empirical {\teff}--color  calibrations  for  solar
metallicity   in   order    to  fix    the   semi-empirical    colors,
$\stackrel{\sim}{c_{ij}}^{emp}$, at a given metallicity:
\begin{eqnarray}
 \lefteqn{       \stackrel{\sim}{c_{ij}}^{emp}\!        \!      \!  \!
 (T,\log g_{\mathrm{seq}},[M/H])            =               c_{ij}^{emp}
 (T,\log g_{\mathrm{seq}},[M/H] \! = \!  0) }  \nonumber \\ & & + \Delta
 c_{ij} (T,\log g_{\mathrm{seq}},\Delta [M/H])\,,
\label{eq:col_semi_emp}
\end{eqnarray}

where $\log g_{\mathrm{seq}}$ designates the surface gravity along the
giant   or  the  dwarf  sequences,  as    defined   by the   empirical
calibrations.      The     resulting    empirical  and  semi-empirical
{\teff}--color calibrations  for  the dwarfs and  giants  are given in
Tables \ref{tab:color_p00}  to  \ref{tab:color_m35}.  Empirical colors
for M dwarfs (2000 K $\sim$ 3500 K)  are given in  the range --3.5 $<$
{\mh} $<$ +0.5,   while for M giants  between  2500 K and  3500  K the
calibrations are defined  for --1.0 $<$  {\mh}  $<$ +0.5.  Theoretical
bolometric corrections, $BC_{V}$ (see Eq.
\ref{eq:bcv}),  are given as derived  from the final grid of corrected
spectra (see  Section  \ref{sect:calib_spec} and   LCB97).  While  for
completeness the semi-empirical calibrations are given for the largest
possible ranges of colors and effective temperatures in Tables
\ref{tab:color_p00}  to \ref{tab:color_m35}, we  should emphasize here
that the UBV magnitudes for  the coolest M dwarf  models -- and  hence
the  corresponding  {\em   differential colors} --   are  still rather
uncertain, as will be shown in Section \ref{sect:corr_spec}.

\section{Calibration of theoretical spectra}
\label{sect:calib_spec}

\subsection{Correction procedure and conservation of the original differential properties}
\label{sect:corr_proc}

The  need to  {\em  re}-calibrate  the  theoretical SEDs  was  clearly
demonstrated in  the previous work,  by comparing (i) the model colors
obtained from the     original synthetic spectra with   the  empirical
temperature--color calibrations, and (ii) the model colors originating
from the different basic libraries between  themselves.  In Fig.  6 of
LCB97,  discrepancies as  large  as $\sim$ 0.5  mag  can be found  for
instance  in U--B and  B--V  for {\teff}  $<$   3500 K.  A  correction
procedure was  developed  in order  to  provide {\em color-calibrated}
theoretical fluxes over a large wavelength range, (typically from U to
L), while  preserving  the  color  differences  due to   gravity   and
metallicity  variations given by  the different  original grids.  This
method was based on the definition  of {\em correction functions} at a
given effective temperature,  $\Phi_{\lambda} (T)$, obtained from  the
ratio of the corresponding  empirical and  theoretical pseudo-continua
at {\em each wavelength}.  The pseudo-continua in turn were calculated
from   the colors and, hence,  the   monochromatic fluxes within  each
wavelength    band as black-bodies   with  smoothed color temperatures
varying with wavelength, $T_{c}(\lambda)$:

\begin{equation}
pc_{\lambda}(T_{\mathrm eff}) \propto B_{\lambda}(T_c(\lambda)).
\label{eq:pc}
\end{equation}

A corrected  spectrum can therefore be  simply computed by multiplying
the original spectrum for a given set  of stellar parameters ({\teff},
{\lg}, {\mh}) with the  corresponding correction function at the  same
effective     temperature.    A   scaling    factor,       $\xi(T,\log
g,\chi)$\footnote{Note  that in the  following we use the more compact
notation   defined  in Paper   I    to designate fundamental   stellar
parameters   by the  equivalence:  $(T,\log g,\chi)\equiv(T_{eff},\log
g,[M/H])$.}, is finally applied to the  corrected spectrum in order to
conserve the   bolometric   flux  (Eq.  \ref{eq:corspec}).     As  the
correction function is, by  definition, a factor  which depends on the
effective temperature  {\em only},      this simple  algorithm    also
preserves, to first  order, the original differential  grid properties
implied by metallicity  and/or    surface gravity changes, and    thus
satisfies the basic requirement imposed on the correction procedure.

While this is true for the monochromatic  magnitude differences -- the
correction   function becomes an  additive  constant  on a logarithmic
scale --, this condition cannot be fully achieved for the (broad-band)
colors,  which represent {\em heterochromatic}  measures  of the flux.
Yet,  if we consider an original  spectrum for a  given set of stellar
parameters,     $S^{o}_{\lambda}(T,\log g_{1},\chi_{1})$,  the magnitude
$m_{i}$ in filter {\em i} is:

\begin{eqnarray}
\lefteqn{m_{i}^{o}(T,\log g_{1},\chi_{1}) = }\nonumber \\ 
 & & -2.5 \log \int_{\lambda_{m_{i}}}^{\lambda_{M_{i}}}
S^{o}_{\lambda}(T,\log g_{1},\chi_{1})P^{i}_{\lambda} \, d\lambda  + K_{i}\, ,
\label{eq:magnitude}
\end{eqnarray}

where $P^{i}_{\lambda}$ is the normalized transmission function of the
filter  $i$  defined between $\lambda_{m_{i}}$  and $\lambda_{M_{i}}$.
For  different values of the gravity  and  the chemical composition at
the same {\teff}, we have

\begin{eqnarray}
\lefteqn{m_{i}^{o}(T,\log g_{2},\chi_{2}) = }\nonumber \\ 
 & & -2.5\log
\int_{\lambda_{m_{i}}}^{\lambda_{M_{i}}}
S^{o}_{\lambda}(T,\log g_{2},\chi_{2})P^{i}_{\lambda} \, d\lambda + K_{i}\, .
\end{eqnarray}

The color difference $\Delta c_{ij}^o$  due to the variations in {\lg}
and $\chi$ are then given by

\begin{eqnarray}
\Delta c_{ij}^{o}&=-2.5\log &
\left( \frac{\displaystyle \int_{\lambda_{m_{i}}}^{\lambda_{M_{i}}} S^{o}_{\lambda}(T,\log g_{1},\chi_{1})P^{i}_{\lambda} d\lambda}{\displaystyle \int_{\lambda_{m_{j}}}^{\lambda_{M_{j}}} S^{o}_{\lambda}(T,\log g_{1},\chi_{1})P^{j}_{\lambda} d\lambda} \right.\nonumber \\
 & & \times \left. \frac{\displaystyle \int_{\lambda_{m_{j}}}^{\lambda_{M_{j}}} S^{o}_{\lambda}(T,\log g_{2},\chi_{2})P^{j}_{\lambda} d\lambda}{\displaystyle \int_{\lambda_{m_{i}}}^{\lambda_{M_{i}}} S^{o}_{\lambda}(T,\log g_{2},\chi_{2})P^{i}_{\lambda} d\lambda} \right).
\end{eqnarray}

Since for the corrected spectra (LCB97),

\begin{equation}
S^{c}_{\lambda}(T,\log g,\chi) = \Phi_{\lambda} (T) \cdot \xi(T,\log g,\chi)
\cdot S^{o}_{\lambda}(T,\log g,\chi),
\label{eq:corspec}
\end{equation}

we have, after elimination of $\xi$ in the color term:

\begin{eqnarray}
\Delta c_{ij}^{c} & =-2.5\log &
\left( \frac{\displaystyle \int_{\lambda_{m_{i}}}^{\lambda_{M_{i}}} \Phi_{\lambda}(T) \, S^{o}_{\lambda}(T,\log g_{1},\chi_{1}) P^{i}_{\lambda} d\lambda}{\displaystyle \int_{\lambda_{m_{j}}}^{\lambda_{M_{j}}} \Phi_{\lambda}(T) \, S^{o}_{\lambda}(T,\log g_{1},\chi_{1}) P^{j}_{\lambda} d\lambda} \right. \nonumber \\
 &\! \! & \! \! \times  \left. \frac{\displaystyle \int_{\lambda_{m_{j}}}^{\lambda_{M_{j}}}  \! \!  \! \! \Phi_{\lambda}(T) \,S^{o}_{\lambda}(T,\log g_{2},\chi_{2}) P^{j}_{\lambda} d\lambda}{\displaystyle \int_{\lambda_{m_{i}}}^{\lambda_{M_{i}}}   \! \!  \! \! \Phi_{\lambda}(T) \, S^{o}_{\lambda}(T,\log g_{2},\chi_{2}) P^{i}_{\lambda} d\lambda} \right)\! \!  .
\label{eq:delta_col}
\end{eqnarray}

Thus the differential colors   are {\em rigorously} preserved by   the
correction procedure ({\ie} $\Delta  c_{ij}^{o} = \Delta  c_{ij}^{c}$)
if $\Phi_{\lambda} (T)$     is  constant between    $\lambda_{m}$  and
$\lambda_{M}$.  In practice, these  rather severe constraints are well
matched -- and the  corresponding $\delta(\Delta c_{ij})$ are small --
if  $\Phi_{\lambda} (T)$  varies   slowly  across the  passbands   or,
equivalently, as long as the filters are not too wide.

Figs. 18 and 19 of  LCB97 show, in fact, that  the color residuals are
negligible over  the   whole  ranges of   UBVRIJHKL   colors and model
parameters.  Significant residuals ($\sim$ 0.05 mag) are found in R--I
at the lowest temperatures because  of the more significant variations
of the correction functions inside the long-tailed R filter.

\subsection{Correction of M dwarf model spectra}
\label{sect:corr_spec}

\begin{figure*}
\epsfxsize=18cm
\epsffile {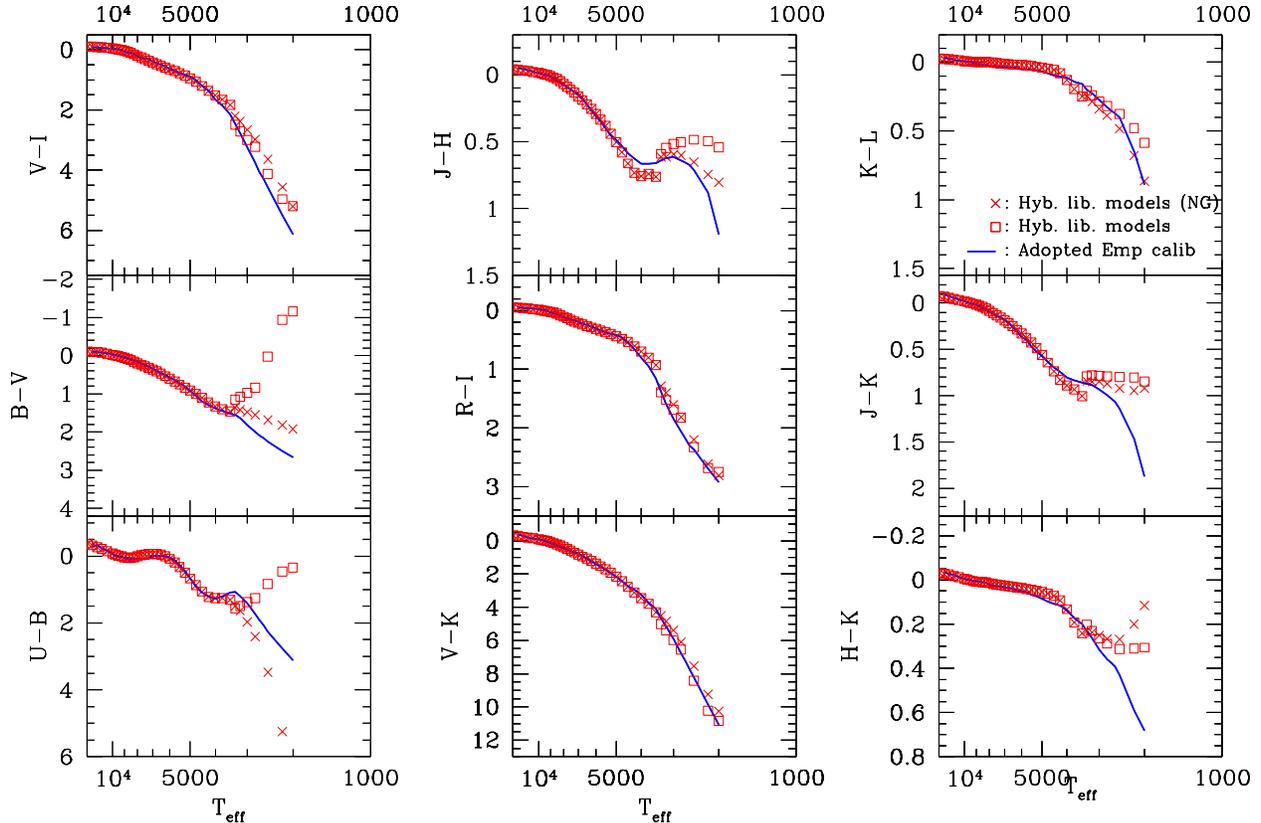}
 \caption []{Colors  of  original models   of  dwarfs compared to  our
 empirical {\teff}--color relations  (solid  lines). The  ``Extended''
 models (open squares) exhibit larger deviations  (except in H--K) and
 unrealistic  U--B  and   B--V colors   compared   to the  ``NextGen''
 (labelled ``NG'') models (crosses).}
\label{f:compar_lbc96_calib_dw.ori}
\end{figure*}

In Fig.  \ref{f:compar_lbc96_calib_dw.ori},  we compare  a sequence of
theoretical colors computed from the dwarf model SEDs to the empirical
colors.  While the hottest models (K95) match  quite well the observed
sequences -- except for {\teff} $<$ 4000  $\sim$ 3750 K -- the coolest
ones  (AH95) exhibit more serious  discrepancies, in particular in the
blue-optical colors.  The {\em Extended} models (open squares) provide
unrealistic U--B and B--V colors, differing  by as much  as 2 $\sim$ 3
mag relative to the {\em NextGen} version models (crosses).  At longer
wavelengths, both these new   and old M  dwarf model  generations seem
more realistic, although the infrared colors appear systematically too
blue    probably   due to an incomplete     $H_{2}O$   opacity list at
solar-metallicity used   in  the  calculations (Allard,  {\em  private
communication}).

\begin{figure*}
\epsfxsize=18cm
\epsffile {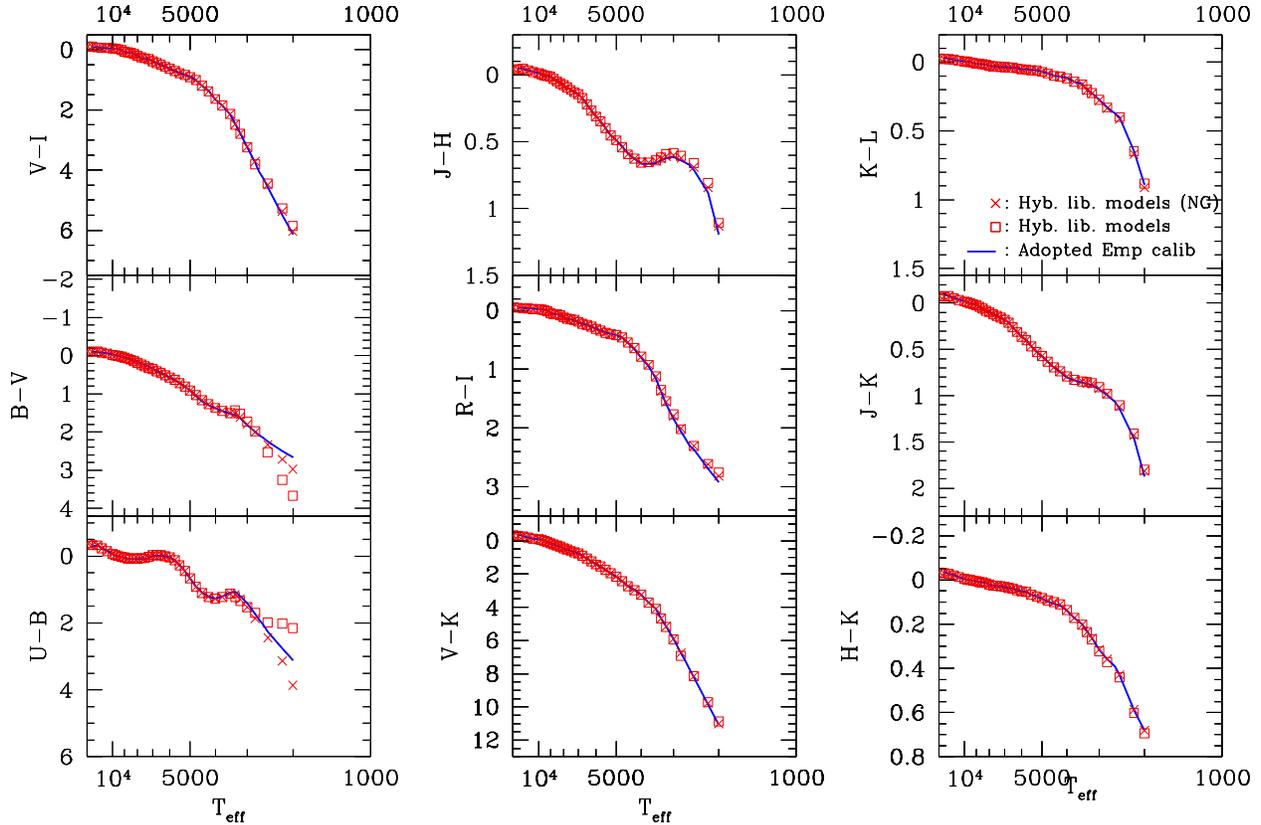}
 \caption []{Same  as  Fig. \ref{f:compar_lbc96_calib_dw.ori}  but for
 the corrected models.  The corrections provide a perfect match of the
 empirical colors over the whole range of  {\teff}, except in U--B and
 B--V below 2500 K, where significant discrepancies persist.}
\label{f:compar_lbc96_calib_dw.cor}
\end{figure*}

In order to provide  more realistic (broad-band)  colors for M dwarfs,
we have  applied   the  correction    method   described  in   Section
\ref{sect:corr_proc}, using the pseudo-continua calculated from  the
updated empirical temperature--color  relationships between 2000 K and
11500 K.    However, in the  range  2000 K  $\sim$ 4500 K   we have to
distinguish  between   the  ``dwarf'' and   the  ``giant''  correction
functions derived from model  spectra originating from different grids
(AH95 and ``B+F'', respectively).  This  was done by fixing the  lower
limit of {\lg}  for  a ``dwarf'' model to    3.0 dex.  Thus,  all  the
spectra with  {\lg} lower than  3.0 and {\teff}  less than  4500 K are
corrected according to the giant empirical  color sequences, while the
others  are calibrated from the  dwarf sequences.  For   {\mh} = 0, we
also defined corrections functions to  be applied uniquely to the {\em
NextGen}   models,   independently of those      computed for the {\em
Extended} models used at other metallicities.  In Fig.
\ref{f:compar_lbc96_calib_dw.cor} we   compare  the  corrected   model
colors to  the  {\teff}--color calibrations: most  of  the theoretical
colors  now match very well  the empirical relations.  For the largest
original deviations found  in  U--B and  B--V below  3000 K, important
discrepancies  still remain (more  than 1  mag  for the {\em Extended}
models), but the corrected  colors  should nevertheless provide   more
reliable  values, in particular those predicted  by  the {\em NextGen}
models.

As for the original  model spectra, UBVRIJHKLM corrected model  colors
and bolometric corrections have been  synthetised for the whole  range
of parameters provided by the complete library.  Color grids are given
in electronic tables accompanying this paper.

\begin{figure*}
\epsfxsize=18cm
\epsffile {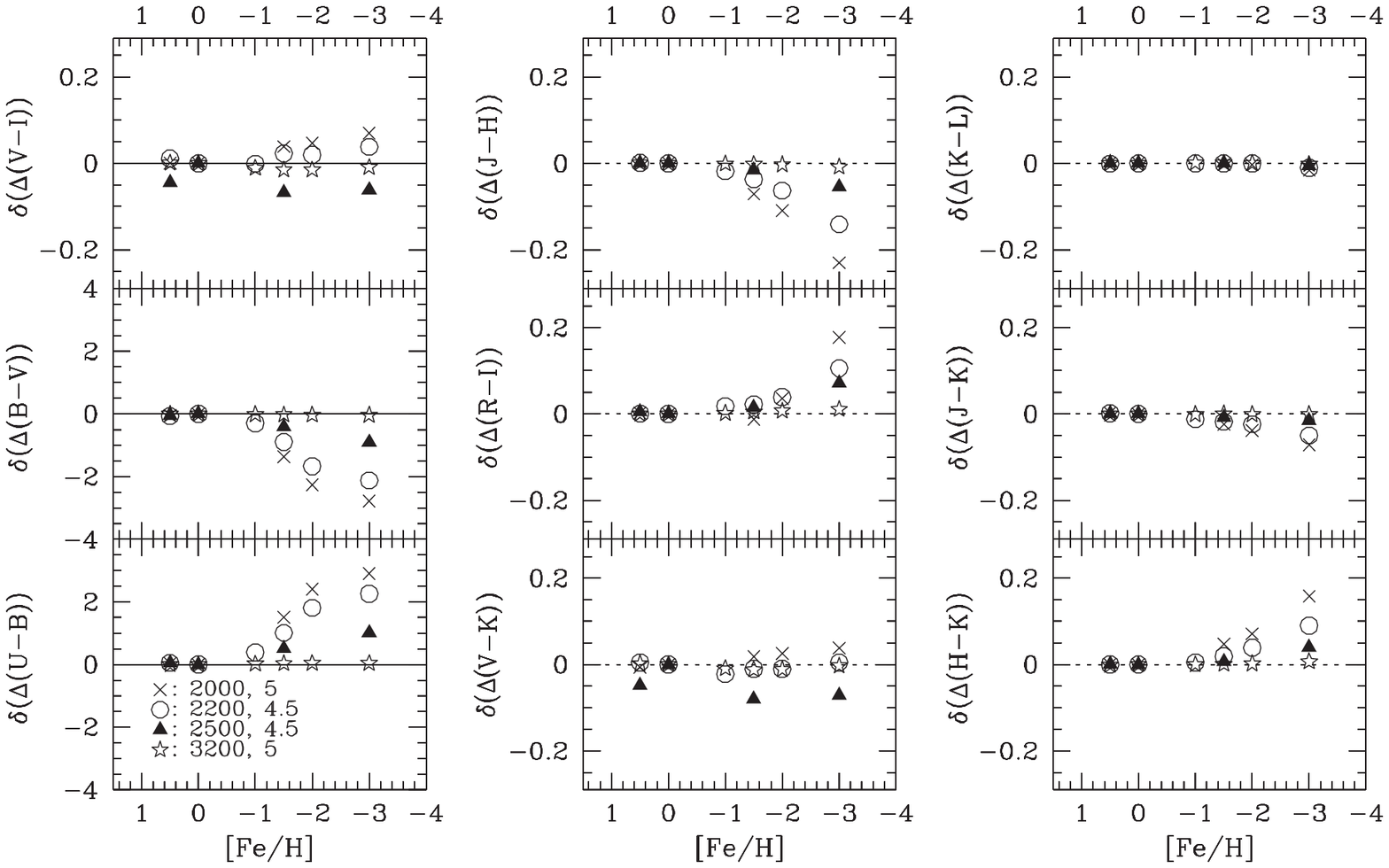}
 \caption     []{Color--difference  residuals  between  corrected  and
 original  models  as functions  of  [Fe/H]  for some of  the  M dwarf
 models. The symbols indicate the  difference between the color excess
 of a corrected model at  a given metallicity and  the color excess of
 the  corresponding original model, having  the same {\teff} and {\lg}
 (as  indicated on the   bottom--left  panel). The color excesses
 are  calculated   as:       $\Delta  c_{ij}  =    c_{ij}(\chi)      -
 c_{ij}(\chi_{\odot})$}
\label{f:diff_effect.Mdw}
\end{figure*}

Because  the  corrections   are so substantial,   the   question which
naturally arises is how  well preserved are the  original differential
colors for the coolest M dwarfs.  We  have computed, for these models,
the residual color  differences between  corrected and original  model
colors,   $\delta(\Delta  c_{ij}) =    (\Delta  c_{ij}^{c}  -   \Delta
c_{ij}^{o})$.  For  metal-content variations the results are presented
in Fig.   \ref{f:diff_effect.Mdw},  where  $\delta(\Delta c_{ij})$  is
plotted as a function of {\mh}.  At low temperatures  ($<$ 2200 K) and
low metallicities ($<$ -2.0), differences as large as $\sim$ 2 mag are
reached for U--B and B--V  !  For the  other colors the residuals  are
smaller, but typical  values of order  $\sim$ 0.2 mag still remain for
the coolest and the most metal-deficient models. Clearly, the original
grid properties are {\em not} conserved for these models.

\begin{figure*}
\epsfxsize=13cm
\centerline{ \epsffile {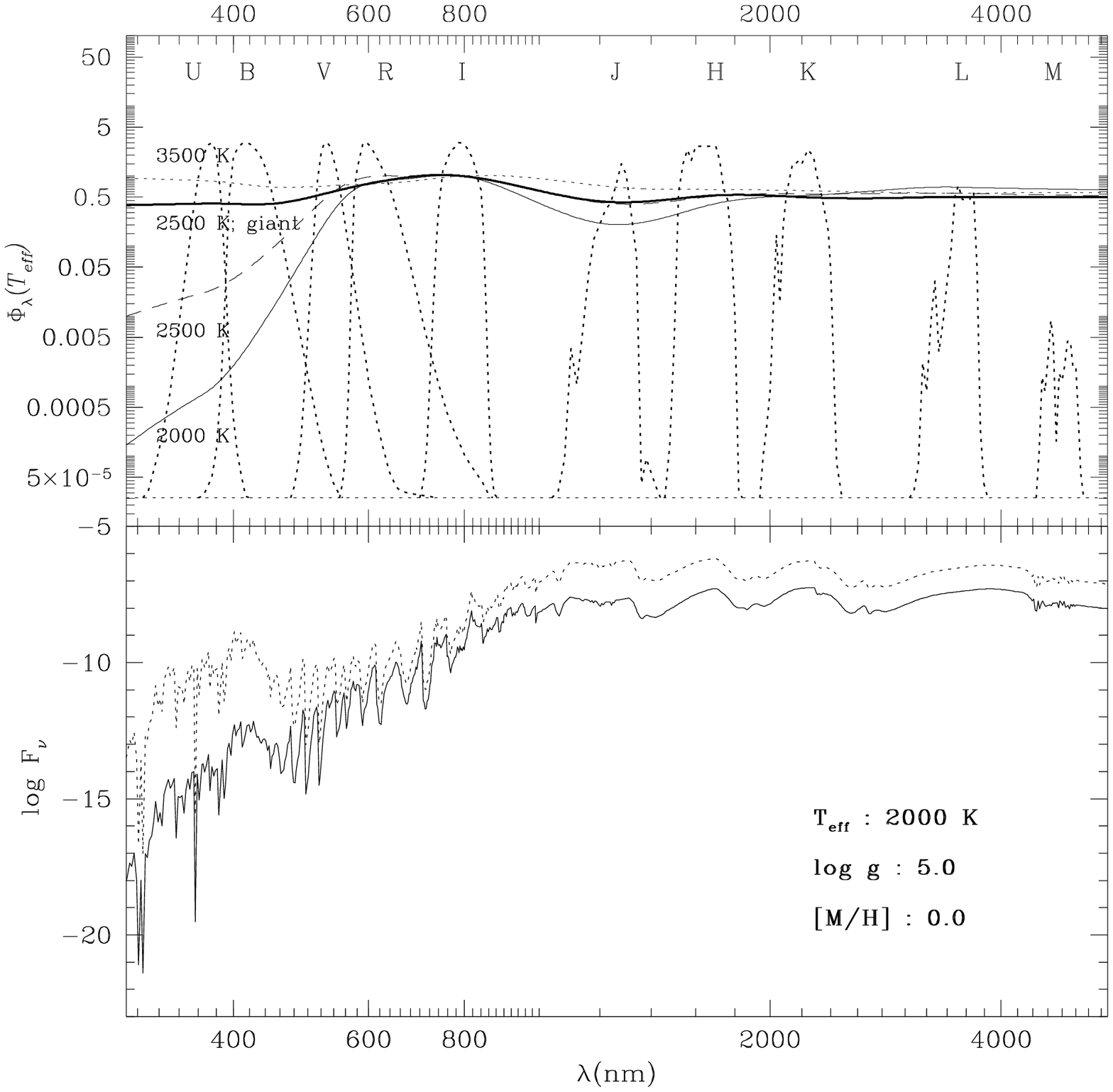}}
 \caption []{Top panel. The correction functions applied to the M star
 models:  for dwarf  spectrum at  2000  K  (thin solid  line),  2500 K
 (long-dashed   line),  3500 K  (short-dashed    line), and for  giant
 spectrum at 2500 K (thick solid line). The different filters are also
 shown. Note the strong variations of these functions in the UBV bands
 for the coolest  dwarf models (2000 K and  2500  K) which affect  the
 differential colors.\\ Bottom  panel.  A  corrected M dwarf  spectrum
 (solid line) compared to the original one (dotted-line). An arbitrary
 shift has been applied for clarity.}
\label{f:corr_fc.Mdw}
\end{figure*}

The main reason behind these deviations is the strong variation of the
correction  factors within the spectral   range covered by  broad-band
filters,   as  discussed     in   Section  \ref{sect:corr_proc}.  Fig.
\ref{f:corr_fc.Mdw} shows the correction  functions computed for  some
of the coolest  models,  compared to  the respective  positions of the
different  passbands, and their effect  on a cool dwarf model spectrum
(bottom panel).  As previously,  the  functions are all normalized  at
817 {nm} in the I band (LCB97)  and plotted on  a logarithmic scale in
order to emphasize the relative differences.  As we can see for 2000 K
(solid line) and 2500  K (short-dashed line), the correction functions
vary considerably across the  UBV filters; {\eg}, the 2000 K--function
changes by a  factor $\sim$ 1000  between 400  and  600 {nm} !   These
steep gradients  inevitably degrade the resulting differential colors.
Significant fluctuations of correction factors are also present in the
JHK bands, accounting for   the  significant residuals seen   in these
differential colors.   For  the giants at   2500 K (thick  line),  the
largest fluctuations appear in   R and I,  leading to  the  deviations
found in LCB97 ($\sim$ 0.05) at this temperature.

\subsection{Shortcomings of the correction algorithm} 
\label{sect:diff_corr}

In order  to investigate correction methods  which better preserve the
differential properties, we   developed a procedure which avoids  the
use of  functions applied  to  the spectra as  multiplicative factors.
This was done  by introducing directly, as an  input to the correction
algorithm,     the      color    differences,    {\nolinebreak $\Delta
c_{ij}(T,\Delta \log g,\Delta \chi)$}, that we want to preserve from the
original  models.   Generalizing the   definitions  of the   non-solar
{\teff}--color  calibrations (Eq.  \ref{eq:col_semi_emp}), we   define
the    semi-empirical    colors of  {\em each}     model  in the grid,
$\stackrel{\sim}{c_{ij}} (T,\log g,\chi)$,  by  adding to the  empirical
colors given in Table \ref{tab:color_p00}  the {\em theoretical} color
differences  due to changes in  metallicity,$\Delta \chi$, and surface
gravity, $\Delta \log g$:

\begin{eqnarray}
 \stackrel{\sim}{c_{ij}}     (T,\log g,\chi)         &       =         &
 c_{ij}^{emp}(T,\log g_{\mathrm{seq}},\chi_{\odot})  \nonumber  \\  &  + & \Delta
 c_{ij}(T,\Delta \log g,\Delta \chi)\,.
\end{eqnarray}

These  semi-empirical colors then    define the {\em   semi-empirical}
pseudo-continuum, $\stackrel{\sim}{pc}_{\lambda} (T,\log g,\chi)$, at  a
given  parameter set in the   grid, following the method described  in
Paper I.   In  order  to preserve  the  detailed information   at  the
resolution   of    the synthetic  spectra,   a  ``spectral function'',
$\Gamma_{\lambda}(T,\log g,\chi)$, obtained by the ratio of the original
model spectrum to the theoretical pseudo-continuum, is then multiplied
with the semi-empirical pseudo continuum, in order  to define the {\em
corrected} spectra (see Appendix):

\begin{equation}
S^{c}_{\lambda}(T,\log g,\chi)      =   \stackrel{\sim}{pc}_{\lambda}
(T,\log g,\chi) * \Gamma_{\lambda}(T,\log g,\chi) \,.
\end{equation}

This ``differential correction'' method  should naturally preserve the
original spectral features and the color differences of the models.

Identical tests as those performed in Section \ref{sect:corr_spec} for
measuring    the   differential   corrected   colors   indicate  that,
unfortunately,   this  ``differential correction''  algorithm fails to
provide significant improvements over the conservation of differential
colors attempted via the previous method.  For models hotter than 3000
K, the   residuals are still  negligible  ($<$ 0.02  mag) and  the two
methods are  really equivalent, but in the  coolest dwarf  regime, the
new algorithm gives even worse results for UBV colors.

\begin{figure*}
\epsfxsize=14cm
\centerline{ \epsffile {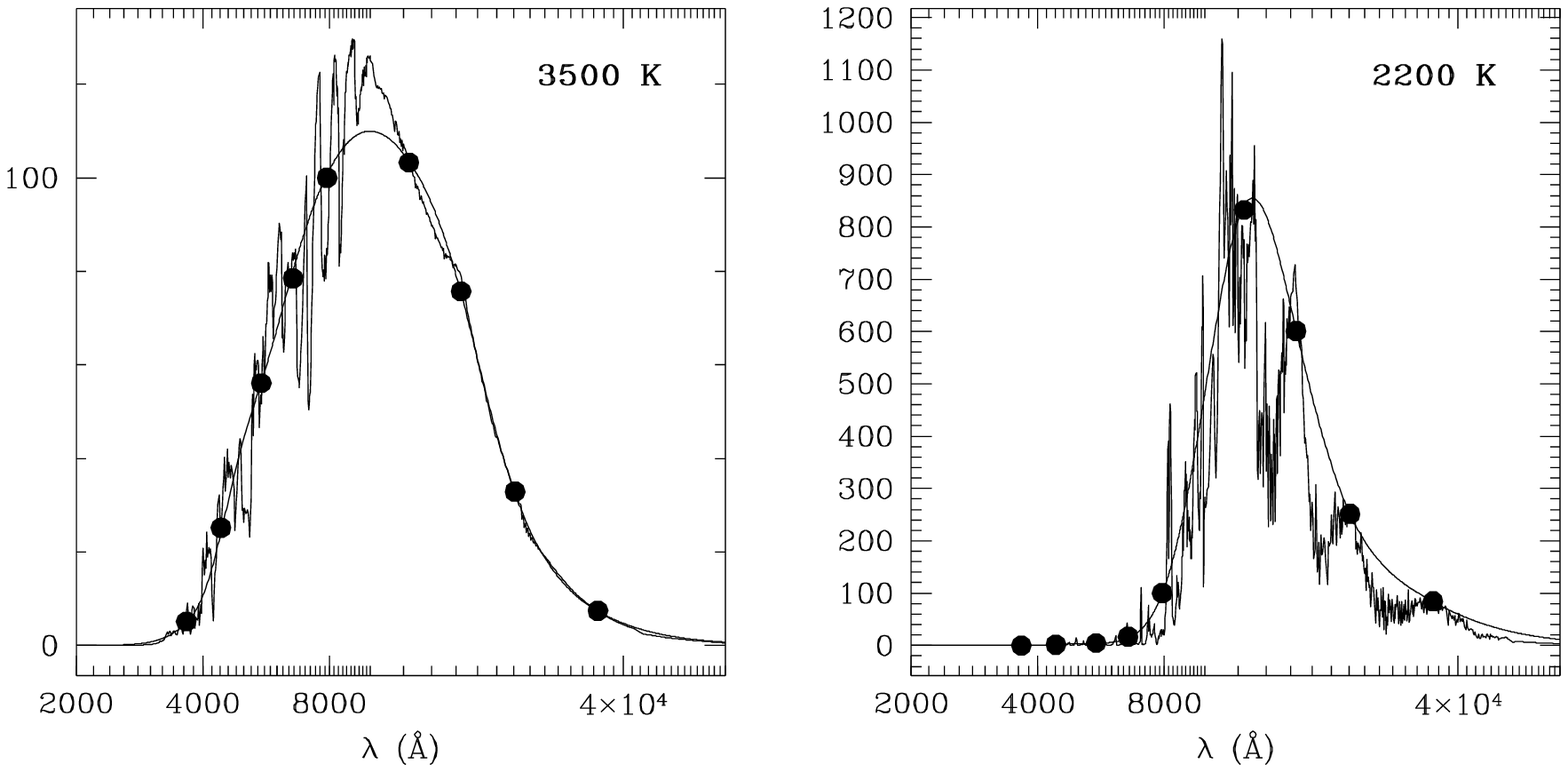}}
 \caption  []{Comparison  of theoretical  spectra  and  their  derived
 pseudo-continua for two values of  {\teff}.  The points indicate  the
 monochromatic fluxes given by the theoretical colors.}
\label{f:compar_spec_pc}
\end{figure*}

Thus,  none  of the two correction  methods  are able to  preserve the
differential color properties for the   coolest star models to  within
the  desired accuracy.  Clearly,  the definition   and use of   a {\em
pseudo-continuum} is inadequate for  such stars. Indeed, at  these low
temperatures, the  presence  of large and strong  molecular absorption
bands    complicates  the stellar  spectra   and   hence also  affects
significantly   the    (broad-band)   colors.   Therefore,    a   {\em
pseudo-continuum}  defined  from  these  colors as    a {\em smoothed}
(black-body) function  cannot  trace the  flux distribution accurately
enough.  As an illustration, the theoretical spectra and their derived
pseudo-continua are compared in Fig.
\ref{f:compar_spec_pc} for two low values of the effective temperature:
for {\teff} = 3500 K (left panel), the pseudo-continuum (normalized in
the K band) follows the  flux variations quite accurately, whereas for
2200 K (right panel), the spectrum is too complex to be described by a
continuous function    such  as a  pseudo-continuum.    Therefore, for
temperatures  less than $\sim$ 2500 K,   a correction function defined
from   smoothed  energy   distributions     is   not    suitable   for
color-calibrating theoretical spectra.  In the future, a more reliable
calibration and correction method  for these complicated spectra
of     the  coolest    stars   must  obviously     be based    on  the
higher--resolution, more detailed observed flux distributions provided
by eight--color  narrow-band photometry (White   \&  Wing 1978) or  by
spectrophotometric data ({\eg} Kirkpatrick {\al} 1991; 1993).

\section{Conclusion}

We have constructed   a comprehensive library  of  theoretical stellar
energy distributions from a  combination  of different basic grids  of
blanketed model atmosphere spectra.   This  new grid  complements  the
preliminary version described in LCB97 by extending it to the M dwarfs
models of Allard \&  Hauschildt (1995).  It provides synthetic stellar
spectra with useful resolution  on a homogeneous wavelength grid, from
9.1 nm to  160  $\mu$m, over  large ranges  of fundamental parameters.
This standard  library should therefore  be  particularly suitable for
spectral evolutionary synthesis  studies of stellar systems and  other
synthetic photometry applications.

Comparison   of  synthetic  photometry with  empirical  {\teff}--color
sequences, established for the  first time down  to 2000 K, have shown
important discrepancies for the coolest  dwarf models.  The correction
procedure   designed   in    the  previous  paper    to  provide  {\em
color-calibrated} fluxes has been extended and applied to the original
dwarf spectra in the  range 4500 K to  2000 K.  Although this seems to
result  in  more realistic    colors, the method   induces significant
changes in the  original  differential color properties.    This stems
from too strongly  variable correction functions  at low temperatures,
resulting from the fact that the pseudo-continuum cannot be adequately
defined for these stars.  The empirical colors for dwarfs below 2500 K
remain uncertain due  to the  lack of  reliable observations.  At  the
lowest temperatures,  the  corrected models should  therefore  be used
with caution.

Despite these   limitations,  we expect    that the corrected  spectra
provide at present  a valuable option  for deriving  realistic stellar
colors   over extensive  ranges   of  temperatures, luminosities,  and
metallicities which are  required  for  reliable population  synthesis
modelling.  Grids of  model spectra and  (UBVRIJHKLM)  colors for both
the original and   the   corrected versions of the    present  stellar
library, as  well  as  the semi-empirical   calibrations  presented in
Tables
\ref{tab:color_p00}  to \ref{tab:color_m35},    are   fully   available
by electronic form at the Strasbourg data center (CDS).

\begin{acknowledgements}
We wish to  thank France Allard and Peter  Hauschildt for making their
extensive  grids of  models  available on  public ftp.   The anonymous
referee is also acknowledged for his helpful  comments.  This work was
supported by the Swiss National  Science Foundation, and this research
has made  use of  the Simbad  database,  operated at CDS,  Strasbourg,
France.
\end{acknowledgements}

\section*{Appendix}

We  give here  a more detailed  description  of the  {\em differential
correction} algorithm discussed in Section \ref{sect:diff_corr}.

For each model  spectrum and given  parameter values $(T,\log g,\chi)$
in the grid, we proceed along the following steps:

\begin{enumerate}
\item
we compute the synthetic colors, $c_{ij}^{o}(T,\log g,\chi)$, from the
original theoretical  energy   distribution,   $S^{o}_{\lambda}(T,\log
g,\chi)$,  and   the normalized  response  functions  of  the filters,
$P^{i}_{\lambda}$, (Eq. \ref{eq:magnitude}) by

\begin{eqnarray}
\lefteqn{c_{ij}^{o}(T,\log g,\chi) = }\nonumber \\
 & & -2.5\log \left( \frac{\displaystyle
\int_{\lambda_{m_{i}}}^{\lambda_{M_{i}}}
S^{o}_{\lambda}(T,\log g,\chi)P^{i}_{\lambda}    d\lambda}{\displaystyle
\int_{\lambda_{m_{j}}}^{\lambda_{M_{j}}}
S^{o}_{\lambda}(T,\log g,\chi)P^{j}_{\lambda} d\lambda} \right)\,;
\end{eqnarray}

\item
as described in detail in  LCB97, these theoretical colors, and  hence
the monochromatic fluxes within each wavelength band, are then used to
calculate           the     {\em   theoretical}      pseudo-continuum,
${pc}^{o}_{\lambda}(T,\log  g,\chi)$,  as a  black-body  with smoothed
color temperatures, $T_{c}(\lambda)$, varying with wavelength:

\begin{equation}
{pc}^{o}_{\lambda}(T,\log g,\chi)                            \propto
B_{\lambda}(T_c(\lambda))\, ;
\label{eq:pc2}
\end{equation}

\item
from the theoretical colors,  $c_{ij}^{o}(T,\log g,\chi)$, we compute the
{\em theoretical color differences},

\begin{eqnarray}
\lefteqn{\Delta  c_{ij}(T,\Delta \log g,\Delta \chi) = }\nonumber \\
&            &              c_{ij}^{o}(T,\log g,\chi)                  -
c_{ij}^{o}(T,\log g_{\mathrm{seq}},\chi_{\odot})\, ,
\end{eqnarray}

where  $\Delta \log g$ and $\Delta  \chi$ are respectively the surface
gravity  and  the metallicity  variations  relative  to the  empirical
sequences --  for dwarfs and giants at  solar metallicity  -- given in
Table
\ref{tab:color_p00}:

\begin{equation}
\Delta \log g = \log g  - \log g_{\mathrm{seq}}\, ,
\end{equation}

with
\[
 \log g_{\mathrm{seq}} = \left\{ \begin{array}{ll}
 \log g_{\, \mathrm{giant}} & \mbox{if $\log g <$ 3.0 and $T \leq$ 4500 K} \\
 \log g_{\, \mathrm{dwarf}} & \mbox{otherwise} \, ,
			      \end{array}
		       \right. 
\]

and

\begin{equation}
\Delta \chi = \chi - \chi_{\odot} \nonumber \,;
\end{equation}

\item
by   adding these  color     differences  to  the empirical    colors,
$c_{ij}^{emp}(T,\log  g_{\mathrm{seq}},\chi_{\odot})$  given in Table
\ref{tab:color_p00}, we then define  the {\em semi-empirical} colors,
$\stackrel{\sim}{c_{ij}} (T,\log g,\chi)$:

\begin{eqnarray}
 \stackrel{\sim}{c_{ij}} (T,\log  g,\chi)  & =   & c_{ij}^{emp}(T,\log
 g_{\mathrm{seq}},\chi_{\odot})   \nonumber    \\  &    +    &  \Delta
 c_{ij}(T,\Delta \log g,\Delta \chi)\,;
\end{eqnarray}    

\item
from Eq. (\ref{eq:pc2}), these {\em  semi-empirical} colors allow us  to
calculate       the      {\em      semi-empirical}   pseudo-continuum,
$\stackrel{\sim}{pc}_{\lambda}(T,\log g,\chi)$;

\item
a ``spectral    function'',   $\Gamma_{\lambda}(T,\log g,\chi)$, which
contains the high--resolution information of the theoretical spectrum,
is defined  by the ratio of the  original spectrum and the theoretical
pseudo-continuum,
 
\begin{equation}
 \Gamma_{\lambda}  (T,\log g,\chi) = S^{o}_{\lambda} (T,\log g,\chi)\:
 / \: pc_{\lambda} (T,\log g,\chi) \,;
\end{equation}

\item
the {\em corrected} spectrum   is finally computed by  multiplying the
spectral function with the semi-empirical pseudo--continuum:

\begin{equation}
S^{c}_{\lambda}(T,\log g,\chi)      =   \stackrel{\sim}{pc}_{\lambda}
(T,\log g,\chi) * \Gamma_{\lambda}(T,\log g,\chi) \,.
\end{equation}

\end{enumerate}

\begin{table*}
\caption{Adopted {\teff}--color relations and {\lg} values for dwarfs and cool giants with {\mh} = 0.0}
\begin{tabular}{rrrrrrrrrrrr}
\hline \hline
 & & & & & & & & & & & \\
\multicolumn{1}{c}{{\teff}} & \multicolumn{1}{c}{U-B} & \multicolumn{1}{c}{B-V} & \multicolumn{1}{c}{V-I} & \multicolumn{1}{c}{V-K} & \multicolumn{1}{c}{R-I} & \multicolumn{1}{c}{J-H} & \multicolumn{1}{c}{H-K} & \multicolumn{1}{c}{J-K} & \multicolumn{1}{c}{K-L} & \multicolumn{1}{c}{$BC_{\mathrm{V}}$} & \multicolumn{1}{c}{{\lg}}\\
 & & & & & & & & & & & \\
\hline
 & & & & & & & & & & & \\
\underline{\em Dwarfs} & & & & & & & & & & & \\
 & & & & & & & & & & & \\
2000. & 3.119 & 2.664 & 6.148 & 11.090 & 2.921 & 1.192 & 0.682 & 1.874 & 0.891 & -7.623 & 5.490\\
2200. & 2.772 & 2.498 & 5.504 & 9.871 & 2.691 & 0.881 & 0.592 & 1.473 & 0.652 & -6.496 & 5.417\\
\hline
\end{tabular}
\label{tab:color_p00}
\end{table*}

\begin{table*}
\caption{Same as Table \ref{tab:color_p00} for {\mh} = +1.0}
\begin{tabular}{rrrrrrrrrrrr}
\hline \hline
 & & & & & & & & & & & \\
\multicolumn{1}{c}{{\teff}} & \multicolumn{1}{c}{U-B} & \multicolumn{1}{c}{B-V} & \multicolumn{1}{c}{V-I} & \multicolumn{1}{c}{V-K} & \multicolumn{1}{c}{R-I} & \multicolumn{1}{c}{J-H} & \multicolumn{1}{c}{H-K} & \multicolumn{1}{c}{J-K} & \multicolumn{1}{c}{K-L} & \multicolumn{1}{c}{$BC_{\mathrm{V}}$} & \multicolumn{1}{c}{{\lg}}\\
 & & & & & & & & & & & \\
\hline
 & & & & & & & & & & & \\
\underline{\em Dwarfs} & & & & & & & & & & & \\
 & & & & & & & & & & & \\
3500. & 0.845 & 1.190 & 2.272 & 4.774 & 1.363 & 0.716 & 0.150 & 0.865 & 0.121 & -2.157 & 4.900\\
3750. & 1.142 & 1.178 & 1.906 & 4.323 & 1.135 & 0.724 & 0.115 & 0.839 & 0.087 & -1.746 & 4.760\\
\hline
\end{tabular}
\label{tab:color_p10}
\end{table*}

\begin{table*}
\caption{Same as Table \ref{tab:color_p00} for {\mh} = +0.5}
\begin{tabular}{rrrrrrrrrrrr}
\hline \hline
 & & & & & & & & & & & \\
\multicolumn{1}{c}{{\teff}} & \multicolumn{1}{c}{U-B} & \multicolumn{1}{c}{B-V} & \multicolumn{1}{c}{V-I} & \multicolumn{1}{c}{V-K} & \multicolumn{1}{c}{R-I} & \multicolumn{1}{c}{J-H} & \multicolumn{1}{c}{H-K} & \multicolumn{1}{c}{J-K} & \multicolumn{1}{c}{K-L} & \multicolumn{1}{c}{$BC_{\mathrm{V}}$} & \multicolumn{1}{c}{{\lg}}\\
 & & & & & & & & & & & \\
\hline
 & & & & & & & & & & & \\
\underline{\em Dwarfs} & & & & & & & & & & & \\
 & & & & & & & & & & & \\
2000. & 3.161 & 2.358 & 6.437 & 11.492 & 2.922 & 1.168 & 0.775 & 1.943 & 0.821 & -7.720 & 5.490\\
2200. & 2.898 & 2.653 & 5.348 & 9.643 & 2.580 & 0.851 & 0.660 & 1.512 & 0.631 & -6.157 & 5.420\\
\hline
\end{tabular}
\label{tab:color_p05}
\end{table*}

\begin{table*}
\caption{Same as Table \ref{tab:color_p00} for {\mh} = -0.5}
\begin{tabular}{rrrrrrrrrrrr}
\hline \hline
 & & & & & & & & & & & \\
\multicolumn{1}{c}{{\teff}} & \multicolumn{1}{c}{U-B} & \multicolumn{1}{c}{B-V} & \multicolumn{1}{c}{V-I} & \multicolumn{1}{c}{V-K} & \multicolumn{1}{c}{R-I} & \multicolumn{1}{c}{J-H} & \multicolumn{1}{c}{H-K} & \multicolumn{1}{c}{J-K} & \multicolumn{1}{c}{K-L} & \multicolumn{1}{c}{$BC_{\mathrm{V}}$} & \multicolumn{1}{c}{{\lg}}\\
 & & & & & & & & & & & \\
\hline
 & & & & & & & & & & & \\
\underline{\em Dwarfs} & & & & & & & & & & & \\
 & & & & & & & & & & & \\
2000. & 2.886 & 2.667 & 6.646 & 11.187 & 3.103 & 1.110 & 0.496 & 1.606 & 1.161 & -7.718 & 5.490\\
2200. & 2.647 & 2.896 & 5.684 & 9.539 & 2.749 & 0.873 & 0.469 & 1.342 & 0.818 & -6.213 & 5.420\\
\hline
\end{tabular}
\label{tab:color_m05}
\end{table*}

\begin{table*}
\caption{Same as Table \ref{tab:color_p00} for {\mh} = -1.0}
\begin{tabular}{rrrrrrrrrrrr}
\hline \hline
 & & & & & & & & & & & \\
\multicolumn{1}{c}{{\teff}} & \multicolumn{1}{c}{U-B} & \multicolumn{1}{c}{B-V} & \multicolumn{1}{c}{V-I} & \multicolumn{1}{c}{V-K} & \multicolumn{1}{c}{R-I} & \multicolumn{1}{c}{J-H} & \multicolumn{1}{c}{H-K} & \multicolumn{1}{c}{J-K} & \multicolumn{1}{c}{K-L} & \multicolumn{1}{c}{$BC_{\mathrm{V}}$} & \multicolumn{1}{c}{{\lg}}\\
 & & & & & & & & & & & \\
\hline
 & & & & & & & & & & & \\
\underline{\em Dwarfs} & & & & & & & & & & & \\
 & & & & & & & & & & & \\
2000. & 2.832 & 3.348 & 6.809 & 9.978 & 3.068 & 0.836 & 0.344 & 1.180 & 1.493 & -6.914 & 5.490\\
2200. & 2.595 & 3.444 & 5.877 & 8.589 & 2.730 & 0.687 & 0.301 & 0.987 & 1.114 & -5.640 & 5.420\\
\hline
\end{tabular}
\label{tab:color_m10}
\end{table*}

\begin{table*}
\caption{Same as Table \ref{tab:color_p00} for {\mh} = -1.5}
\begin{tabular}{rrrrrrrrrrrr}
\hline \hline
 & & & & & & & & & & & \\
\multicolumn{1}{c}{{\teff}} & \multicolumn{1}{c}{U-B} & \multicolumn{1}{c}{B-V} & \multicolumn{1}{c}{V-I} & \multicolumn{1}{c}{V-K} & \multicolumn{1}{c}{R-I} & \multicolumn{1}{c}{J-H} & \multicolumn{1}{c}{H-K} & \multicolumn{1}{c}{J-K} & \multicolumn{1}{c}{K-L} & \multicolumn{1}{c}{$BC_{\mathrm{V}}$} & \multicolumn{1}{c}{{\lg}}\\
 & & & & & & & & & & & \\
\hline
 & & & & & & & & & & & \\
\underline{\em Dwarfs} & & & & & & & & & & & \\
 & & & & & & & & & & & \\
2000. & 3.066 & 4.562 & 6.175 & 7.786 & 2.855 & 0.570 & 0.251 & 0.821 & 1.786 & -5.202 & 5.490\\
2200. & 2.822 & 4.486 & 5.252 & 6.627 & 2.535 & 0.396 & 0.187 & 0.583 & 1.417 & -4.176 & 5.420\\
\hline
\end{tabular}
\label{tab:color_m15}
\end{table*}

\begin{table*}
\caption{Same as Table \ref{tab:color_p00} for {\mh} = -2.0}
\begin{tabular}{rrrrrrrrrrrr}
\hline \hline
 & & & & & & & & & & & \\
\multicolumn{1}{c}{{\teff}} & \multicolumn{1}{c}{U-B} & \multicolumn{1}{c}{B-V} & \multicolumn{1}{c}{V-I} & \multicolumn{1}{c}{V-K} & \multicolumn{1}{c}{R-I} & \multicolumn{1}{c}{J-H} & \multicolumn{1}{c}{H-K} & \multicolumn{1}{c}{J-K} & \multicolumn{1}{c}{K-L} & \multicolumn{1}{c}{$BC_{\mathrm{V}}$} & \multicolumn{1}{c}{{\lg}}\\
 & & & & & & & & & & & \\
\hline
 & & & & & & & & & & & \\
\underline{\em Dwarfs} & & & & & & & & & & & \\
 & & & & & & & & & & & \\
2000. & 3.453 & 5.601 & 5.126 & 5.341 & 2.497 & 0.436 & 0.141 & 0.577 & 2.102 & -3.346 & 5.490\\
2200. & 3.258 & 5.349 & 4.299 & 4.526 & 2.165 & 0.192 & 0.114 & 0.306 & 1.685 & -2.601 & 5.420\\
\hline
\end{tabular}
\label{tab:color_m20}
\end{table*}

\begin{table*}
\caption{Same as Table \ref{tab:color_p00} for {\mh} = -2.5}
\begin{tabular}{rrrrrrrrrrrr}
\hline \hline
 & & & & & & & & & & & \\
\multicolumn{1}{c}{{\teff}} & \multicolumn{1}{c}{U-B} & \multicolumn{1}{c}{B-V} & \multicolumn{1}{c}{V-I} & \multicolumn{1}{c}{V-K} & \multicolumn{1}{c}{R-I} & \multicolumn{1}{c}{J-H} & \multicolumn{1}{c}{H-K} & \multicolumn{1}{c}{J-K} & \multicolumn{1}{c}{K-L} & \multicolumn{1}{c}{$BC_{\mathrm{V}}$} & \multicolumn{1}{c}{{\lg}}\\
 & & & & & & & & & & & \\
\hline
 & & & & & & & & & & & \\
\underline{\em Dwarfs} & & & & & & & & & & & \\
 & & & & & & & & & & & \\
2000. & 4.018 & 6.123 & 4.269 & 3.501 & 2.077 & 0.389 & 0.059 & 0.447 & 2.401 & -2.135 & 5.490\\
2200. & 3.596 & 5.693 & 3.630 & 2.897 & 1.807 & 0.071 & 0.050 & 0.121 & 1.965 & -1.610 & 5.420\\
\hline
\end{tabular}
\label{tab:color_m25}
\end{table*}

\begin{table*}
\caption{Same as Table \ref{tab:color_p00} for {\mh} = -3.0}
\begin{tabular}{rrrrrrrrrrrr}
\hline \hline
 & & & & & & & & & & & \\
\multicolumn{1}{c}{{\teff}} & \multicolumn{1}{c}{U-B} & \multicolumn{1}{c}{B-V} & \multicolumn{1}{c}{V-I} & \multicolumn{1}{c}{V-K} & \multicolumn{1}{c}{R-I} & \multicolumn{1}{c}{J-H} & \multicolumn{1}{c}{H-K} & \multicolumn{1}{c}{J-K} & \multicolumn{1}{c}{K-L} & \multicolumn{1}{c}{$BC_{\mathrm{V}}$} & \multicolumn{1}{c}{{\lg}}\\
 & & & & & & & & & & & \\
\hline
 & & & & & & & & & & & \\
\underline{\em Dwarfs} & & & & & & & & & & & \\
 & & & & & & & & & & & \\
2000. & 4.278 & 6.211 & 3.657 & 1.634 & 1.701 & 0.163 & -0.215 & -0.052 & 2.960 & -1.365 & 5.490\\
2200. & 3.766 & 5.724 & 3.154 & 1.433 & 1.495 & -0.119 & -0.110 & -0.229 & 2.353 & -1.026 & 5.420\\
\hline
\end{tabular}
\label{tab:color_m30}
\end{table*}

\begin{table*}
\caption{Same as Table \ref{tab:color_p00} for {\mh} = -3.5}
\begin{tabular}{rrrrrrrrrrrr}
\hline \hline
 & & & & & & & & & & & \\
\multicolumn{1}{c}{{\teff}} & \multicolumn{1}{c}{U-B} & \multicolumn{1}{c}{B-V} & \multicolumn{1}{c}{V-I} & \multicolumn{1}{c}{V-K} & \multicolumn{1}{c}{R-I} & \multicolumn{1}{c}{J-H} & \multicolumn{1}{c}{H-K} & \multicolumn{1}{c}{J-K} & \multicolumn{1}{c}{K-L} & \multicolumn{1}{c}{$BC_{\mathrm{V}}$} & \multicolumn{1}{c}{{\lg}}\\
 & & & & & & & & & & & \\
\hline
 & & & & & & & & & & & \\
\underline{\em Dwarfs} & & & & & & & & & & & \\
 & & & & & & & & & & & \\
2000. & 4.436 & 6.149 & 3.177 & 0.446 & 1.334 & 0.191 & -0.423 & -0.232 & 3.378 & -0.929 & 5.490\\
2200. & 3.889 & 5.671 & 2.787 & 0.548 & 1.199 & -0.066 & -0.270 & -0.336 & 2.639 & -0.695 & 5.420\\
\hline
\end{tabular}
\label{tab:color_m35}
\end{table*}

\end{document}